\documentclass[titlepage,12pt]{article}
\usepackage{amsmath,amssymb}
\begin{document}

\begin{titlepage}
\thispagestyle{empty}
\begin{center}
{\Large\bf Wave packets in quantum theory of collisions}\\[0.5cm]
{\large\bf M. I. Shirokov }\\[0.5cm]
{Bogoliubov Laboratory of Theoretical Physics\\
Joint Institute for Nuclear Research\\
141980 Dubna, Russia\\
e-mail: shirokov@theor.jinr.ru}
\end{center}

\vspace*{1cm} \noindent
{\large Abstract}\\%[0.5cm]

Two methodological troubles of the quantum theory of collisions are considered.
The first is the undesirable interference of the incident and scattered waves
in the stationary approach to scattering. The second concerns the nonstationary
approach to the theory of collisions of the type $a+b\rightarrow c+d$.
In order to calculate the cross section one uses the matrix element
$\langle cd|S|ab\rangle$ of the $S$-matrix. The element is proportional
to $\delta$-function expressing the energy conservation. The corresponding
probability $|\langle cd|S|ab\rangle|^2$ contains $\delta^2$ which is
mathematically senseless. The known regular way to overcome the difficulty
seems to be unsatisfactory. In this paper, both the troubles are resolved using
wave packets of incident particles.
\end{titlepage}

%%%%%%%%%%%%%%%%%%%%%%%%%%%%%%%%%%%%%%%%%%%%%%%%%%%%%%%%%%%%%%%%%%%%%%%%%%%

\section{INTRODUCTION}
\label{intro}

An approach to the theory of scattering is known which may be called stationary.
The scattering is considered as a stationary process: there is a steady flux
of particles incident on a potential $V$ (the target). Scattered particles
are also described by a steady flux. The state of the system is described by
a vector which is constant in time. The vector is an eigenstate of the total
Hamiltonian $H=H_0+V$ which belongs to the continuous spectrum of $H$:
$H\psi_k=E_k\psi_k$. The eigenstate is known to be the superposition
of the {\bf i}ncident wave $I_{\vec{k}}(\vec{x})$ and {\bf s}cattered wave
$S_{\vec{k}}(\vec{x})$:
\begin{equation}
\label{eq1}
\Psi_{\vec{k}}(\vec{x})=I_{\vec{k}}(\vec{x})+
S_{\vec{k}}(\vec{x}),\quad I_{\vec{k}}(\vec{x})={\rm e}^{i\vec{k}\vec{x}},
\quad S_{\vec{k}}(\vec{x})=A(\vartheta,\varphi){\rm e}^{ikr}/r.
\end{equation}
Here $k=\sqrt{2mE_k}$ and $\vec{x}=(r,\vartheta,\varphi)$, $\vartheta$
being the angle between $\vec{k}$ and $\vec{x}$. The axis $z$ is chosen to be
parallel to the momentum $\vec{k}$ of the incident particle ($\vec{x}$
being its position).

The stationary approach uses the probability flux or its density
\begin{equation}
\label{eq2}
\vec{j}(\vec{x})=\frac{i}{2m}\left[ \vec{\nabla}\psi^*(\vec{x})
\psi (\vec{x})-\psi^*(\vec{x})\vec{\nabla}\psi (\vec{x}) \right]
\end{equation}
instead of the usual probability $|\psi (\vec{x})|^2$, e.g., see
\cite{3}, Ch.~IV, \S~29; \cite{5}, Ch.~II, \S~15, Ch.~XI, \S~95;
\cite{10}, Ch.~II.7.
The probability flux generated by $\Psi_{\vec{k}}=I_{\vec{k}}+
S_{\vec{k}}$ (see Eq.~(\ref{eq1})) is
\begin{eqnarray}
\label{eq3}
\vec{j}_{\vec{k}}&=&\frac{i}{2m}\left( \vec{\nabla}I_{\vec{k}}^*\cdot
I_{\vec{k}} -c.c. \right) +\frac{i}{2m}\left( \vec{\nabla}S_{\vec{k}}^*\cdot
S_{\vec{k}} -c.c. \right)\nonumber\\
&+&\frac{i}{2m}\left( \vec{\nabla}S_{\vec{k}}^*\cdot
I_{\vec{k}} -c.c. \right) +\frac{i}{2m}\left( \vec{\nabla}I_{\vec{k}}^*\cdot
S_{\vec{k}} -c.c. \right).
\end{eqnarray}

The cross section is determined by the ratio of the scattered flux (the second term
in Eq.~(\ref{eq3})) to the incident one (the first term in Eq.~(\ref{eq3})), see
Sect.~\ref{S2} below. Besides these fluxes the total flux $\vec{j}_{\vec{k}}$
contains the interference terms (the third and fourth in Eq.~(\ref{eq3})). Their
physical sense is unclear. It is unknown what contribution the interference
terms may bring in the cross section. One may conjecture that they must vanish
if one replaces the plane incident wave $I_{\vec{k}}$ by an incident wave packet,
see \cite{9}, Ch.~X.5; \cite{10}, Ch.~V, end of \S~18.
This conjecture is confirmed in Sect.~\ref{S3}. The wave
packet is used which tends to the plane wave $I_{\vec{k}}=\exp (i\vec{k}\vec{x})$
when packet dimension increases. Other ways of the packet introduction are possible.
For example, Messiah~\cite{9} used a classical ensemble of small packets which have
different impact parameters. Finally, averaging over the ensemble is carried out,
see \cite{9}, Ch.~X.

The stationary theory of scattering is nonrelativistic and is inapplicable, e.g.,
to the photon scattering (photon position operator and density of flux have no
satisfactory definitions). The nonstationary approach is applicable to any process
of the type $a+b\rightarrow c+d$. It is based on the solution of the Schroedinger
equation for the operator $U(t,t_0)$ of evolution in time. Initially, at the
moment $t_0$, the system is in a state $\Psi(t_0)=|ab\rangle$. At the moment $t$
the system state is described by the vector $\Psi(t)=U(t,t_0)\Psi(t_0)$. The
probability to find a final state $|cd\rangle$ at the moment $t$ is equal to
$|\langle cd|U(t,t_0)|ab\rangle|^2$. The limit $t_0\rightarrow -\infty$,
$t\rightarrow +\infty$, i.e. $S$-matrix, is usually considered.

The approach has the following trouble. The matrix element $\langle cd|S|ab\rangle$
is known to be proportional to the $\delta$-function which expresses the total
energy conservation: the total initial energy $E_a+E_b$ is equal to the final
energy $E_c+E_d$
\begin{equation}
\label{eq4}
\langle cd|S|ab\rangle\sim\delta (E),\quad E=(E_c+E_d)-(E_a+E_b).
\end{equation}
The probability $|\langle cd|S|ab\rangle|^2$ is proportional to
the square $\delta^2$ of this $\delta$-function. This quantity
does not exist mathematically, see~\cite{1}. Physicists gave to
$\delta^2$ an interpretation, see the end of Sect.~\ref{S4}, but
it cannot be recognized as satisfactory. Another resolution of
this trouble is known, e.g., see \cite{7}, Ch.~I4; \cite{10},
Ch.~VIII. It is presented in Sect.~\ref{S4} using the packet
description of the initial state $|ab\rangle$.

So a packet description of the incident particle allows us to resolve two
troubles of the collision theory stated above.

%%%%%%%%%%%%%%%%%%%%%%%%%%%%%%%%%%%%%%%%%%%%%%%%%%%%%%%%%%%%%%%%%%%%%%%%%%%%%%%%%

\section{Definitions of cross section}
\label{S2}

The density of the incident flux $F$ is defined as the number of incident particles
crossing per unit time a unit surface placed perpendicular to the direction of
propagation. Let $\rho$ be the number of particles per unit volume and $\vec{v}$
be the velocity of the incident particles. Then $\vec{F}=\rho\vec{v}$. If there is
one particle in 1 cm$^3$, then $\vec{F}=\vec{v}$.

Let us assume that the coordinate origin is in the center of a target. Let $j_r$
be the density of the probability flux of the scattered particles. The probability
(or the number of particles) going through the area element $rd\varphi r\sin \vartheta
d\vartheta\equiv r^2d\Omega$ during one unit of time is equal to $j_r r^2d\Omega$.
This is the probability $\Delta N$ to detect the particle in the solid angle
$\Delta\Omega$ during one unit of time (e.g., one second)
\begin{equation}
\label{eq5}
\Delta N = j_r r^2\Delta\Omega .
\end{equation}
The quantity $\Delta N$ may be related to the probability $\Delta W(t)$ of the
particle detection in the solid angle at the moment $t$. One may assume
\begin{equation}
\label{eq6}
\Delta N = \Delta W(t+1\, {\rm sec})-\Delta W(t)\cong\frac{d}{dt}\Delta W(t)
\cdot 1\, {\rm sec}.
\end{equation}
Usually another relation is assumed
\begin{equation}
\label{eq7}
\Delta N = W(t)/t.
\end{equation}
Relations (\ref{eq6}) and (\ref{eq7}) coincide if time derivative
$d\Delta W(t)/dt$ is constant. One has the relation
\begin{equation}
\label{eq8}
j_r r^2\Delta\Omega = \Delta N = \Delta W(t)/t.
\end{equation}
Therefore, the definition
\begin{equation}
\label{eq9}
\Delta\sigma = \frac{j_r}{F}r^2 d\Omega
\end{equation}
of the cross section (see \cite{3}, Ch.~XIII; \cite{5}, Ch.~XI, \S~95) is
equivalent to
\begin{equation}
\label{eq10}
\Delta\sigma = \Delta N/F,
\end{equation}
cf. \cite{9}, Ch.~10.

%%%%%%%%%%%%%%%%%%%%%%%%%%%%%%%%%%%%%%%%%%%%%%%%%%%%%%%%%%%%%%%%%%%%%%%%%%%%%%%%%%%%%%%

\section{Packets in stationary approach}
\label{S3}

There is the conjecture that the description of an incident particle by a packet
(instead of the plane wave) will turn into zero interference terms in the total flux
(the third and fourth terms in Eq.~(\ref{eq3})). Let us give a confirmation of this
conjecture.

The introduction of a packet implies that scattering is no longer a stationary process.
Even if the scattering potential is absent, the (free) packet shifts and spreads.

One possible setting of the problem of a packet scattering will be
considered here (for another approach, see, e.g., \cite{9},
Ch.~X). The definition of the cross section as the ratio of fluxes
(see Eq.~(\ref{eq9})) will be retained though fluxes will not be
stationary. The natural requirement is assumed (and ensured): in
the limit when the packet turns into a plane wave the result should
go to the usual stationary one.

Consider a superposition
\begin{equation}
\label{eq11}
\Psi (\vec{x})=\int d^3k \tilde{I}_{\vec{k}} \Psi_{\vec{k}}(\vec{x})
\end{equation}
of the $H$ eigenfunctions $H\Psi_{\vec{k}}=E_k\Psi_{\vec{k}}$. For $\Psi_{\vec{k}}$
see (\ref{eq1}), for $\tilde{I}_{\vec{k}}$ see below. The superposition is not $H$
eigenfunction, but the vector
\begin{equation}
\label{eq12}
\Psi (\vec{x},t)=\int d^3k {\rm e}^{-iE_kt} \Psi_{\vec{k}} (\vec{x})
I_{\vec{k}}
\end{equation}
is a solution of the equation $\partial\Psi (t)/\partial t=H\Psi (t)$.

The vector $\Psi (\vec{x},t)$ consists of two parts
\begin{eqnarray}
\label{eq13}
\Psi (\vec{x},t)&=& I(\vec{x},t)+S(\vec{x},t),\\
\label{eq14}
I(\vec{x},t)&=&\int d^3k {\rm e}^{-iE_kt} \tilde{I}_{\vec{k}} {\rm e}^{i\vec{k}\vec{x}},\\
\label{eq15}
S(\vec{x},t)&=&\int d^3k {\rm e}^{-iE_kt} \tilde{I}_{\vec{k}} A(\vartheta,\varphi)
{\rm e}^{ikr}/r.
\end{eqnarray}
The vector $I(\vec{x},t)$ is the known description of the moving free
packet (it is assumed that the spectrum of $H=H_0+V$ is the same
as the $H_0$ spectrum). In order to calculate
$S(\vec{x},t)$ and the cross section, the following program is accepted.\\
$(a)$ The initial wave packet $I(\vec{x},0)$ will be chosen.\\
$(b)$ This determines the coefficients $\tilde{I}_{\vec{k}}$ in Eqs.~(\ref{eq11}),
(\ref{eq15}) and, therefore, $S(\vec{x},t)$ may be calculated.\\
$(c)$ Absence of the interference terms will be verified.\\
$(d)$ Incident and scattering fluxes may then be found as well as the cross section.

$(a)$ Choice of $I(\vec{x},0)$.

Consider the auxiliary wave function $f(\vec{x})$ which is concentrated in a ball of
the radius $R_I$ ($V_I$ being the ball volume). The ball will be named
``support of $f(\vec{x})$".

\underline{Note.} In other words ``support" is defined here as the volume outside
which the function practically vanishes (or is unobservably small). In mathematics
a different definition is accepted: the support is the volume outside which the
function is exactly zero.

If $f(\vec{x})$ is spherically symmetric, then the average position $\int d^3x \vec{x}
f^*(\vec{x})f(\vec{x})$ is zero. Fourier transform of $f(\vec{x})$ is also spherically
symmetric and, therefore, the average momentum also equals zero. The average
position of the shifted function $f(\vec{x}-\vec{a})\equiv I_{\vec{a}}(\vec{x})$
is equal to $\vec{a}$, average momentum being zero as before. One may verify that
the function ${\rm e}^{i\vec{p}\vec{x}}f(\vec{x})\equiv I_{\vec{p}}(\vec{x})$ has
average momentum $\vec{p}$. At last, consider the shifted function
$I_{\vec{p}}(\vec{x})$, i.e., the function
\begin{equation}
\label{eq16}
I_{\vec{p}}(\vec{x}-\vec{a})={\rm e}^{i\vec{p}(\vec{x}-\vec{a})}f(\vec{x}-\vec{a})
\equiv I_{\vec{p}\vec{a}}(\vec{x}).
\end{equation}
Its average position is $\vec{a}$ and average momentum is $\vec{p}$ (the factor
$\exp (-i\vec{p}\vec{a})$ may be omitted). Let $\tilde{I}_{\vec{p}\vec{a}}(\vec{k})$
be the Fourier transform of $I_{\vec{p}\vec{a}}(\vec{x})$.

Let us assume
\begin{equation}
\label{eq17}
f(\vec{x})=\left\{
\begin{array}{l}
  0,\,\, \mbox{outside the ball}\,\, V_I \\
  1,\,\, {\rm inside}\,\, V_I
\end{array}
\right.
\end{equation}
This means that $I_{\vec{p}}(\vec{x})=f(\vec{x})\exp (i\vec{p}\vec{x})$ is equal to
$\exp (i\vec{p}\vec{x})$ inside $V_I$. When $R_I\rightarrow\infty$
the vector $I_{\vec{p}}(\vec{x})$
tends to the plane wave whose wave function is equal to $\exp (i\vec{p}\vec{x})$
everywhere. We have $\int d^3x |I_{\vec{p}}(\vec{x})|^2=V_I$: there is one particle
in the unit volume.

The initial (at the moment $t=0$) wave function of the incident packet is chosen to
be equal to
\begin{equation}
\label{eq18}
I_{\vec{p}\vec{a}}(\vec{x})={\rm e}^{i\vec{p}\vec{x}}f(\vec{x}-\vec{a}).
\end{equation}
The packet center $\vec{a}$ is placed at the point $(-R_I)$ on the
$z$ axis which is parallel to $\vec{p}$ and passes through the
potential center. This choice means that the interaction of the
packet with the potential begins at the moment $t=0$ and stops at
the moment $T=2R_I/v$, $v=P/m_0$ (the packet dimension $R_I$ is
assumed to be much greater than the dimension $R_v$ of the volume in
which the potential is concentrated). Till the moment $t=0$ and
after the moment $T$ the potential does not act on the packet and
it is free.

$(b)$ Choosing $I_{\vec{p}\vec{a}}(\vec{x})$ one may calculate $S(\vec{x},t)$
(see (\ref{eq15}))
using the Fourier transform $\tilde{f}(\vec{p}-\vec{k})\exp (-i\vec{k}\vec{a})$
of $I_{\vec{p}\vec{a}}(\vec{x})$. If the packet $I_{\vec{p}\vec{a}}(\vec{x})$ has
a macroscopical dimension (e.g., 1 cm), then $\tilde{f}(\vec{p}-\vec{k})$ has a sharp
maximum at $\vec{k}\cong\vec{p}$. Therefore, $E_k\cong E_p$ and $|\vec{k}|\cong |\vec{p}|$.
Let us calculate $S(\vec{x},t)$ approximately using the expansions of $E_k$ and
$|\vec{k}|$ in Taylor series about the point $\vec{p}$:
\begin{eqnarray}
\label{eq19}
E_k&\cong &E_p+(\vec{k}-\vec{p})\vec{v},\quad |\vec{k}|\cong |\vec{p}|
+(\vec{k}-\vec{p})\vec{u},\\
\label{eq20}
\vec{v}&=&\vec{p}/m,\quad \vec{u}=\vec{p}/p,\quad \vec{v}=\vec{u}v.
\end{eqnarray}
The scattering amplitude
$A(\vartheta_k,\varphi_k)$ is simply replaced by
$A(\vartheta_p,\varphi_p)$, $\vartheta_k$ and $\vartheta_p$ being
the angles between $\vec{x}$ and $\vec{k}$, $\vec{p}$,
respectively, cf. Eq.~(\ref{eq1}). One obtains
\begin{equation}
\label{eq21}
S(\vec{x},t)=A(\vartheta_p,\varphi_p)\frac{1}{r}\exp i(pr-E_pt)
\int d^3k {\rm e}^{i(\vec{k}-\vec{p})(\vec{u}r-\vec{v}t)}
{\rm e}^{-i\vec{k}\vec{a}}\tilde{f}(\vec{p}-\vec{k}).
\end{equation}
After the change $\vec{k}'=\vec{p}-\vec{k}$ of variables in $\int d^3k\ldots$
one gets
\begin{eqnarray}
S(\vec{x},t)&=&A\frac{1}{r}\exp i(pr-E_pt){\rm e}^{-i\vec{p}\vec{a}}
\int d^3k' \tilde{f}(\vec{k}'){\rm e}^{i(\vec{k}'\vec{b})},\nonumber\\
\vec{b}&\equiv&-\vec{u}(r-vt)+\vec{a},\quad \vec{a}=-\vec{u}R_I.
\nonumber
\end{eqnarray}
Here $\int d^3k' \tilde{f}(\vec{k}')\exp (i\vec{k}'\vec{b})$ is Fourier
representation of $f(\vec{b})$. Therefore
\begin{equation}
\label{eq22}
S(\vec{x},t)=A\frac{1}{r}\exp i(pr-E_pt){\rm e}^{-i\vec{p}\vec{a}}
f(-\vec{u}[(r-vt)+R_I]).
\end{equation}
Remind that $f(\vec{b})=1$ when $|\vec{b}|\leq R_I$, i.e.
$|-\vec{u}[(r-vt)+R_I]|\leq R_I$ or
\begin{equation}
\label{eq23}
|r-vt+R_I|\leq R_I,
\end{equation}
because $|\vec{u}|=1$ and $f$ is spherically symmetric. The support of
$S(\vec{x},t)$ is determined by inequality (\ref{eq23}). Let us
discuss it.

One may assume that $r\gg R_I$: the packet dimension $R_I$ is much less
than the distance between the target and the detector. Then (\ref{eq23})
may be replaced by a simpler inequality
\begin{equation}
\label{eq24}
|r-vt|\leq R_I.
\end{equation}
It determines the movement of the scattered packet. Let us compare it with
the movement of the initial incident packet $I_{\vec{p}}(\vec{x})=
f(\vec{x})\exp i\vec{p}\vec{x}$, see Eq.~(\ref{eq18}). The latter is the
shift $\vec{x}\rightarrow\vec{x}-\vec{v}t$ along the $z$ axis, if the spreading
is neglected, (e.g., see \cite{12}, Ch.~10.4):
\begin{equation}
\label{eq25}
I_{\vec{p}}(\vec{x},t)=I_{\vec{p}}(\vec{x}-\vec{v}t,0)
=f(\vec{x}-\vec{v}t)\exp i\vec{p}(\vec{x}-\vec{v}t), \quad \vec{v}=\vec{p}/m.
\end{equation}
This means the shift of the packet support, i.e., the ball of the
radius $R_I$ (the phase factor $\exp i\vec{p}\vec{v}t$ is
inessential). The shifted ball is described by the inequality
\begin{equation}
\label{eq26}
|\vec{x}-\vec{v}t|\leq R_I.
\end{equation}
The shifted ball is not a spherically symmetric region (excluding
the case when the packet centre coincides with the coordinate
origin, i.e., potential centre). Meanwhile, inequality
(\ref{eq24}) describes a spherically symmetric region at all times:
(\ref{eq24}) does not contain angles $\vartheta$, $\varphi$ of the
vector $\vec{x}$ but contains only $|\vec{x}|=r$. At fixed $t$
the support region is the spherical layer between the spheres of
the radii $vt-R_I$ and $vt+R_I$. The thickness of the layer is
equal to $2R_I$. As $t$ increases, this layer ``inflates"
preserving its thickness.

When $t$ is small and $|r-vt|\cong |r|$, then inequality (\ref{eq23})
is not fulfilled and $S(\vec{x},t)=0$: the scattered wave appears in the
detector only when $t$ is sufficiently large.

$(c)$ We obtain that the scattered packet is in the spherical layer described above,
while the support of the incident packet moves along the $z$ axis. The moving packets'
supports do not practically intersect if $r$ is sufficiently large.
Therefore, $S(\vec{x},t)I_{\vec{p}\vec{a}}(\vec{x},t)=0$: $S$ is zero where $I$
is nonzero and vice versa. So the interference of the incident and scattered
waves is absent (with the exception of small values of the angle $\vartheta$).

$(d)$ The used expressions (\ref{eq25}) and (\ref{eq22}) for incident and
scattered wave packets differ from the corresponding $I_k$ and $S_k$
waves in Eq.~(\ref{eq1}) only in one respect: the former have an additional
factor $f$ which is equal to unity inside the moving packets and vanishes
outside them. Therefore, the incident and scattered fluxes \underline{inside}
packets are the same as in the stationary case. However, these fluxes are nonstationary:
their supports move in space. One may retain the previous definition (\ref{eq9})
for the cross section having in mind that the fluxes $F$ and $j_v$ in (\ref{eq9})
are intrapacket ones.

Note once more that the used packet's description of scattering turns into
the ordinary stationary one when $R_I\rightarrow\infty$ (the condition
$r\gg R_I$ being implied).

%%%%%%%%%%%%%%%%%%%%%%%%%%%%%%%%%%%%%%%%%%%%%%%%%%%%%%%%%%%%%%%%%%%%%%%%%%%%%%%%%%%%%%

\section{Packets in nonstationary approach}
\label{S4}

The nonstationary approach to the collision theory uses the evolution
operator $U(t,t_0)$ (interaction or Dirac picture is in mind). For the
reaction $a+b\rightarrow c+d$ one must calculate the matrix
element of the type $\langle cd|U(t,t_0)|ab\rangle$. Let us assume
that the initial state $|i\rangle =|ab\rangle$ is the product of
packets $|a\rangle$, $|b\rangle$, see Sect.~\ref{S3}. For example
$$
|a\rangle =\int d^3k |\vec{k}\rangle a(\vec{k}).
$$
The packets have finite supports in the coordinate space. The
supports are supposed to be of macroscopically large dimensions
and, therefore, packet spreading may be neglected, see \cite{6},
Ch.~3.1; \cite{12}, Ch.~10.4.

The particles do not interact if their supports do not intersect.
So the interaction lasts during a finite interval $T$ of time. In
the following I let $t_0=-T/2$, $t=T/2$.

Let us consider the matrix element of the expansion of
$U(T/2,-T/2)$ in the perturbation series
\begin{eqnarray}
\label{eq27}
&&U_{fi}(T)\equiv \langle f|U(T/2,-T/2)|i\rangle
=\langle f|i\rangle +i\int_{-T/2}^{T/2} dt {\rm e}^{i(E_f-E_i)t}
\langle f|H^s_{int}|i\rangle \nonumber\\
&&+ i^2\sum_m \int_{-T/2}^{T/2} dt_1 {\rm e}^{i(E_f-E_m)t_1}
\int_{-T/2}^{t_1} dt_2 {\rm e}^{i(E_m-E_i)t_2} \langle
f|H^s_{int}|m\rangle \langle m|H^s_{int}|i\rangle +\ldots
\end{eqnarray}
Cf. \cite{8}, Chs.~1.2 and 1.3; \cite{11}, Ch.~11.6. Here $E_f$
denotes the final total energy $E_f\equiv E_c+E_d$. Analogously,
$E_i\equiv E_a+E_b$, ($E_a$ and $E_b$ being average energies);
$E_m$ is the total energy of the intermediate state $|m\rangle$;
$H^s_{int}$ is the interaction Hamiltonian in the Schroedinger
picture. One may suppose that $\langle f|i\rangle =0$.

The second term in Eq.~(\ref{eq27}) is proportional to
\begin{equation}
\label{eq28}
2\pi\delta_T (E)\equiv \int_{-T/2}^{T/2} dt
{\rm e}^{iEt}=\frac{2\sin ET/2}{E}, \quad E\equiv E_f-E_i.
\end{equation}
The third term contains the integral over $t_2$
\begin{equation}
\label{eq29}
\int_{-T/2}^{t_1} dt_2 {\rm e}^{i(E_m-E_i)t_2}
=\left[ {\rm e}^{i(E_m-E_i)t_1}
-{\rm e}^{i(E_m-E_i)(-T/2)}\right]/i(E_m-E_i).
\end{equation}
The contribution to (\ref{eq29}) originating from the lower limit
$-T/2$ tends to zero as $T\rightarrow\infty$ due to fast
oscillations of $\exp i(E_m-E_i)T/2$. For a strict proof of this
statement one must use the packet description of $|i\rangle$ and
Riemann--Lebesque lemma, see, e.g., \cite{11}, Ch.~11 after
Eq.~(11.165). Neglecting this contribution one obtains that the
remaining integral over $t_1$ is equal to $2\pi\delta_T (E)$,
Eq.~(\ref{eq28}). Analogously, one may argue that in all orders of
the expansion (\ref{eq27}) $U_{fi}(T)$ is proportional to
$\delta_T (E_f-E_i)$
\begin{equation}
\label{eq30}
U_{fi}(T)\cong\delta_T (E_f-E_i)\langle f|R|i\rangle,
\end{equation}
where $\langle f|R|i\rangle$ ceases to depend upon $T$ as
$T\rightarrow\infty$. Note that $\delta_T (E)\rightarrow
\delta (E)$ as $T\rightarrow\infty$. In this limit
Eq.~(\ref{eq27}) turns into
\begin{equation}
\label{eq31}
S_{fi}=2\pi\delta (E_f-E_i)\langle f|R|i\rangle,
\end{equation}
where $S$ is the $S$-matrix. One gets that the probability
$|\langle f|U(\infty,-\infty)|i\rangle|^2$ is proportional to
$\delta^2$.
The square of the $\delta$-function has no mathematical sense, see
\cite{1}, part~III, Sect.~12.5.

However, for the cross section we need probability in unit of time,
see Sect.~\ref{S2}. It may be defined as $|U_{fi}(T)|^2/T$. It
follows from Eqs.~(\ref{eq30}) and (\ref{eq28}) that
\begin{equation}
\label{eq32}
|U_{fi}(T)|^2/T \sim \frac{4\sin^2 ET/2}{TE^2}.
\end{equation}
The r.h.s. of Eq.~(\ref{eq32}) tends to $2\pi\delta (E)$ as
$T\rightarrow\infty$, not to $\delta^2 (E)$, see \cite{7}, Ch.~2,
Eq.~(8.19). So we obtain the following value for the probability
in unit time:
\begin{equation}
\label{eq33}
\lim_{T\rightarrow\infty} |U_{fi}(T)|^2/T = 2\pi\delta (E_f-E_i)
|\langle f|R|i\rangle|^2.
\end{equation}

The probability in unit time may be defined in a different way, namely
as $d|U_{fi}(T)|^2/dT$. We have
\begin{equation}
\label{eq34}
\frac{d}{dT} |U_{fi}(T)|^2 \sim \frac{d}{dT}
\frac{4\sin^2 ET/2}{E^2} = \frac{2}{E} \sin ET/2.
\end{equation}
In the limit $T\rightarrow\infty$ one gets in the r.h.s. of
Eq.~(\ref{eq33}) the $\delta (E)$ function as above, cf.
\cite{11}, Ch.~11, Eq.~(11.91).

The usual way to calculate the cross section is to start with
the $S$-matrix element $\langle f|U(\infty,-\infty)|i\rangle$. The
reason is that it is relativistic invariant, allows
renormalization etc., unlike $\langle f|U(t,t_0)|i\rangle$. But
the probability $|\langle f|S|i\rangle|^2$ is proportional to
$\delta^2 (E_f-E_i)$ and this is senseless. The trouble is usually
overcome in the following manner (see, e.g., \cite{8},
Ch.~1.2; \cite{11}, Ch.~I4.1; \cite{2}, Ch.~5, \S~37; \cite{4}).
In the product $\delta (E)\delta (E)$ one of the $\delta$-functions
is replaced by $\delta (0)$ because of the presence of the other
$\delta$-function. Basing on the representation
\begin{equation}
\label{eq35}
\delta (E)=\lim_{T\rightarrow\infty}
\frac{1}{2\pi}\int_{-T/2}^{T/2} {\rm e}^{iEt} dt
\end{equation}
$\delta (0)$ is replaced by $T/2\pi$. In order to obtain the
probability in unit of time one divides $|\langle f|S|i\rangle|^2$
by $T$. So one gets
$$
|\langle f|S|i\rangle|^2/T \sim \delta (E)
$$
which is a sensible result.

However, the argumentation is not satisfactory: $\delta (E)$ does not
depend on $T$ unlike $\delta_T (E)$, see Eq.~(\ref{eq28}) ($\delta
(E)$ is the limit of (\ref{eq28}) as $T\rightarrow\infty$).
Instead of (\ref{eq35}) one may use the representation
$$
\delta (E)=\lim_{T\rightarrow\infty} \frac{1}{2\pi}\int_{-T}^{T}
{\rm e}^{iEt} dt
$$
and analogously obtain $\delta (0)=2T/2\pi$ instead of
$\delta (0)=T/2\pi$.

Nevertheless, the resulting ``probability in unit time" obtained in
the books cited above coincides with the r.h.s. of
Eq.~(\ref{eq33}) obtained here.

%%%%%%%%%%%%%%%%%%%%%%%%%%%%%%%%%%%%%%%%%%%%%%%%%%%%%%%%%%%%%%%%%%%%%%%%%%%%%%%%%%%%%%

\section{CONCLUSION}
\label{S5}

As has been expected, the interference of the incident and
scattered waves in the stationary theory of scattering is absent
if the waves are described by packets. The cause is the
nonintersection of the packets' supports (excepting the limitingly
small scattering angles).

Consideration of the $\delta^2$-trouble arising in the
nonstationary approach needs the determination of the
``probability in unit time" $W_{fi}(T)$. In Sect.~\ref{S4}, for
this purpose the evolution operator $U(T/2,-T/2)$ was used,
$W_{fi}(T)$ being defined as
\begin{equation}
\label{eq36}
W_{fi}(T)=|\langle f|U(T/2,-T/2)|i\rangle|^2/T.
\end{equation}
In the limit $T\rightarrow\infty$ one gets (see Sect.~\ref{S4}),
\begin{equation}
\label{eq37}
\lim W_{fi}(T)\sim \delta (E),\quad E=E_f-E_i.
\end{equation}
The result is free of the $\delta^2$ trouble.

Another way to get ``probability in unit time" is given in books
on quantum field theory, e.g., see \cite{8}, \cite{11}, \cite{2},
\cite{4}. At first, one considers the $\lim U(T/2,-T/2)$ as
$T\rightarrow\infty$, i.e., the $S$-matrix (in our way the limit
$T\rightarrow\infty$ is carried out later, see (\ref{eq36}),
(\ref{eq37})). The $S$-matrix elements $\langle f|S|i\rangle$ are
proportional to $\delta (E)$. The corresponding probability
$|\langle f|S|i\rangle|^2$ is proportional to $\delta^2 (E)$, which
is senseless. In the books $\delta^2$ is treated in an
unsatisfactory manner presented and criticized at the end of
Sect.~\ref{S4}.

Both the ways give the same result, Eq.~(\ref{eq33}). Here the
satisfactory way of getting the result is considered.

%%%%%%%%%%%%%%%%%%%%%%%%%%%%%%%%%%%%%%%%%%%%%%%%%%%%%%%%%%%%%%%%%%%

%\newpage
%\vspace*{1cm} \noindent
%{\large\bf REFERENCES}%\\[0.5cm]
%\begin{description}
%\item
%\end{description}

%%%%%%%%%%%%%%%%%%%%%%%%%%%%%%%%%%%%%%%%%%%%%%%%%%%%%%%%%%%%%%%%%%%%%%%%%%%%%%%%%%%%%%

\end{document}